\documentclass[preprint]{aastex}

\shorttitle{\ion{O}{6} in the Scutum Supershell}
\shortauthors{Sterling et al.}

\begin{document}

\title{FUSE Observations of \ion{O}{6} Overlying the Scutum Supershell}

\author{N. C. Sterling\altaffilmark{1}, B. D. Savage\altaffilmark{1}, P. Richter\altaffilmark{1}, D. Fabian\altaffilmark{1}, and K. R. Sembach\altaffilmark{2}}

\altaffiltext{1}{Astronomy Department, University of Wisconsin, 475 North Charter Street, Madison, WI 53706}
\altaffiltext{2}{Department of Physics and Astronomy, Johns Hopkins University, 3400 N. Charles Street, Baltimore, MD 21218}

\begin{abstract}

We present \emph{Far Ultraviolet Spectroscopic Explorer} observations of the B0~III star HD~177989 ($l = 17.8^{\rm o}, b = -11.9^{\rm o}, d$ = 4.9 kpc).  This line of sight passes through the high latitude outflow from the Scutum Supershell (GS~018-04+44), a structure that lies at a kinematic distance of $\sim$~3.5 kpc, and spans $\sim 5^{\rm o}$ in diameter.  The \ion{O}{6} $\lambda$1031.93 line is compared with STIS observations of \ion{Si}{4}, \ion{C}{4}, and \ion{N}{5} to examine the ionizing mechanisms responsible for producing the high ion absorption, as well as to study the processes by which gas is expelled into the halo.  The \ion{O}{6} profile spans a similar velocity range as the other highly ionized atoms, from $-70$ to +80 km~s$^{-1}$.  Component fits reveal very broad absorption at the kinematic velocity of the Scutum shell, which differs from the narrow \ion{Si}{4}, \ion{C}{4}, and \ion{N}{5} components, suggesting that these other species occupy a more confined region.  It is possible that the \ion{O}{6} is present in hot gas from the shell interior that is diffusing into the high latitude region above it.  The column densities in the Scutum Supershell component are N(\ion{Si}{4}) = $(3.59 \pm 0.09) \times 10^{13}$, N(\ion{C}{4}) = $(1.78 \pm 0.04) \times 10^{14}$, N(\ion{N}{5}) = $(8.89 \pm 0.79) \times 10^{12}$, and N(\ion{O}{6}) = $(7.76 \pm 0.75) \times 10^{13}$ cm$^{-2}$.  The corresponding column density ratios of N(\ion{C}{4})/N(\ion{Si}{4}) = 4.96 $\pm$ 0.17, N(\ion{C}{4})/N(\ion{N}{5}) = 20.0 $\pm$ 1.8, and N(\ion{C}{4})/N(\ion{O}{6}) = 2.29 $\pm$ 0.23 show that while the \ion{C}{4} and \ion{Si}{4} columns are amplified in this region, the enhancement is not reflected in \ion{N}{5} or \ion{O}{6}.  We suggest that such ionic ratios and column densities could be produced by $\sim$~150--200 turbulent mixing layers lying in a fragmented medium above the polar cap of the superbubble.  We note through a sight line comparison that although these absorption characteristics are similar to those near the center of Radio Loops I and IV, they differ considerably from those near the edges of the loops.  It is apparent that the traits of high ion absorption in a supershell, as well as the ionizing mechanisms responsible, can vary among sight lines through the shell.

\end{abstract}


\section{INTRODUCTION}

The majority of supernovae occur in or near the plane of the Galaxy.  When a series of these transpire in an OB association, their energy combines with winds from massive stars to carve a supershell out of the interstellar medium (Tenorio-Tagle \& Bodenheimer 1988).  If these processes are sufficiently energetic, the shell may break through the material of the Galactic disk and jettison hot gas into the lower density halo (Mac Low et al. 1989).  Absorption-line spectroscopy of high ionization species such as \ion{C}{4}, \ion{N}{5}, and \ion{O}{6} allow astronomers to study the hot matter in these regions, and provide a key to understanding the physical state of such shells and the mechanisms by which gas is fed into the halo.

Past studies of supershells by means of absorption-line spectroscopy have revealed unusual ionization conditions.  For example, Sembach, Savage, \& Tripp (1997) report that gas within Radio Loops I and IV is enriched in \ion{C}{4} 2--3 times compared to adjacent lines of sight that pass outside these remnants.  Similarly, \ion{C}{4} is enhanced significantly in the Scutum Supershell (Savage, Sembach, \& Howk 2001, hereafter SSH).  The ratios of N(\ion{C}{4})/N(\ion{N}{5}) = 10.7 $\pm$ 2.1 and 20.0 $\pm$ 1.8 for HD~119608 (Radio Loop IV) and HD~177989 (Scutum Supershell) are far above the Galactic average of 4.6 $\pm$ 2.4 (Sembach et al. 1997), and rank among the largest observed in the Milky Way.

The \emph{Far Ultraviolet Spectroscopic Explorer} (\emph{FUSE}) satellite provides access to the far ultraviolet (FUV) wavelength region between 905 and 1187 \AA.  With the exceptions of short-lived observing programs conducted with the Hopkins Ultraviolet Telescope (HUT) and the Orbiting and Retrievable Far and Extreme Ultraviolet Spectrometer (ORFEUS), this wavelength regime has not been accessible since the days of the \emph{Copernicus} satellite (1972--1981).  The \ion{O}{6} $\lambda\lambda$1032, 1038 doublet in the FUV is a sensitive tracer of the hot component of the ISM, which is present in shock-heated supershells.  \ion{O}{6} is produced primarily by collisional ionization rather than photoionization, since the ionization potential of \ion{O}{5} (113.9 eV) lies far above the strong photospheric \ion{He}{2} absorption edge (54.4 eV) observed in the spectra of hot stars.  By contrast, \ion{Si}{3} and \ion{C}{3} have ionization potentials of 33.5 and 47.9 eV.  Thus, \ion{Si}{4} and \ion{C}{4} are somewhat more ambiguously identified with hot interstellar gas than \ion{O}{6}, since it is possible to produce them by photoionization.  In collisional ionization equilibrium, \ion{O}{6}, \ion{N}{5}, and \ion{C}{4} peak in abundance at $T \approx $(3, 2, 1)$\times 10^{5}$ K, respectively (Sutherland \& Dopita 1993).  However, hot gas at these temperatures cools rapidly and the assumption of equilibrium ionization may be invalid.  Therefore, theories of the origin of these ions often involve non-equilibrium ionization processes (Spitzer 1990, 1996).

In this paper we present a measurement of interstellar \ion{O}{6} absorption in the spectrum of the star HD~177989 obtained by the \emph{FUSE} satellite.  This line of sight samples gas in a high latitude cloud that is believed to be associated with the Scutum Supershell (Callaway, et al. 2000).  The primary focus of this paper is to relate the \ion{O}{6} measurement with those of \ion{Si}{4}, \ion{C}{4}, and \ion{N}{5} from SSH in order to study the ionization conditions and the responsible processes.

This paper is structured as follows:  The HD~177989 sight line and properties of the Scutum Supershell are reviewed in \S2.  In \S3 we discuss the reduction of the data and the methods of analysis.  The observational results are presented in \S4, followed by a discussion of the \ion{O}{6} absorption and the salient ionization processes that may produce the observed ionic ratios in \S5.  In \S6 we compare our observations to the sight line of QSO~3C~273, which lies in the direction of Radio Loops I and IV.  We provide a summary of the principal results of our study in \S7.

\section{THE HD~177989 SIGHT LINE AND THE SCUTUM SUPERSHELL}

Studies of both the HD~177989 sight line (Savage, Sembach, \& Howk 2001; SSH) and the Scutum Supershell (Callaway et al. 2000) have appeared recently in the literature.  We briefly summarize some of the pertinent information about the sight line and supershell presented in these papers.

HD~177989 is an inner-galaxy B0~III star located in the direction $l = 17.8^{\rm o}, b = -11.9^{\rm o}$.  Its spectroscopic distance is calculated to be 4.9 kpc, based upon its colors ($V = 9.33, B-V = -0.05$; Hill et al. 1974), the color excess $E(B-V) = 0.25$, and the relation $A_{\rm V} = 3.1E(B-V)$ for total visual extinction.  We assumed an absolute magnitude of $-4.90$ for a B0 III star (Walborn 1972) and an intrinsic $B-V$ color of $-0.30$ from Johnson (1963).  The distance of 4.9 kpc places HD~177989 at 1.0 kpc below the disk and at a Galactocentric distance of 4.3 kpc.  Using the Clemens (1985) Galactic rotation curve, the presumed radial velocity of gas in this direction relative to the Local Standard of Rest (LSR) should increase by approximately 11 km~s$^{-1}$ kpc$^{-1}$ for the first 3 kpc, and reach a velocity of +65 km~s$^{-1}$ at the distance of HD~177989.  In the direction of HD~177989, $v_{\rm LSR} - v_{\rm Helio} = +10.5$ km~s$^{-1}$, where the LSR correction is calculated for a solar motion of 16.5 km~s$^{-1}$ in the direction $l = 53^{\rm o}, b = +25^{\rm o}$ (see Mihalas \& Binney 1981).  All velocities given in this paper are referenced to the LSR.

This line of sight penetrates gas from the Local, Sagittarius, and Scutum spiral arms.  Assuming distances of $\sim 2$ and 3--4 kpc for the latter two arms, the sight line samples gas lying respectively $\sim 0.4$ and 0.6--0.8 kpc below the Galactic plane.  Corotating halo gas under these arms is expected to have radial velocities of $\sim +20$ and +30 to +50 km~s$^{-1}$, respectively (Callaway et al. 2000).  This paper focuses mainly on interstellar absorption in the Scutum Arm, where the supershell is located.

The Scutum supershell (GS $018-04+44$) is seen in \ion{H}{1} 21~cm emission as an elongated shell of about 5 degrees in diameter, extending to $\sim 7$ degrees below the plane of the Galaxy.  The bright edges of the shell surround a spherical region containing a noticeable dearth of \ion{H}{1} 21~cm emission, which is most evident at a velocity of +44 km~s$^{-1}$.  This velocity corresponds to a kinematic distance of 3--4 kpc; at 3.5 kpc, the above angular dimensions of the supershell imply a diameter of $\sim 300$ pc.  The top boundary of the shell ($b < -5^{\rm o}$) is absent, but a cloud of \ion{H}{1} that is most conspicuous at +44 km~s$^{-1}$ exists at even higher latitudes (extending to $b = -12^{\rm o}$, or $z = -670$ pc).  The similarity in radial velocities of the high latitude \ion{H}{1} cloud and the supershell is highly suggestive of a blowout event, where energetic processes near the Galactic plane heat and eject hot disk gas into the lower density halo.  ROSAT X-ray observations lend credence to the conjecture of supershell blowout in this region (Snowden et al. 1995).  After correcting for foreground X-ray absorption, Callaway et al. (2000) find that X-ray emission at 1.5 keV peaks near the base of the shell, 0.75 keV inside and at the top of the shell, and 0.25 keV towards latitudes above the shell.  This corroborates the idea that hot gas from inside the shell is lifted to higher latitudes, cooling with increasing distance from its source.

HD~177989 lies behind the high latitude cloud associated with the supershell.  SSH present high resolution spectra of this star obtained with the Space Telescope Imaging Spectrograph (STIS) and the Goddard High Resolution Spectrograph (GHRS).  They find that absorption in the Scutum Arm from the high ions \ion{Si}{4}, \ion{C}{4}, and \ion{N}{5} is at velocities intermediate to the double peaked absorption seen in neutral and low ions such as \ion{N}{1}, \ion{S}{2}, \ion{Si}{2}, and \ion{Fe}{2}.  More specifically, the high ions absorb near +42 km~s$^{-1}$, while the low ions absorb at $\sim +37$ and +55 km~s$^{-1}$.  This finding suggests that the low ions trace the swept up gas bounding the region, while the higher ions trace the hot gas inside the shell.  We provide additional comments on the STIS and GHRS spectra in \S4.

\section{OBSERVATIONS AND DATA REDUCTION}

The \emph{FUSE} design comprises four co-aligned prime focus telescopes and Rowland-circle spectrographs with microchannel plate detectors.  Two channels have Al:LiF coatings, which provide optimal throughput between 1000 and 1187 \AA, while the other two channels have SiC coatings for optimal reflectivity below 1000 \AA.  Moos et al. (2000) and Sahnow et al. (2000) provide a more detailed account of the \emph{FUSE} design and performance.  Twenty exposures of HD~177989 were obtained on 28 August 2000, for a total exposure time of 10,289 seconds.  The data were collected with the large (LWRS, $30\farcs 0 \times 30\farcs 0$) aperture in the spectral image (histogram) mode.  The \emph{FUSE} data are archived at the Multi-Mission Archive at the Space Telescope Science Institute (MAST) under the data set identification P1017101.

The raw data were processed with the CALFUSE (Vers. 1.8.7) standard reduction pipeline available at the Johns Hopkins University as of December 2000.  A total of 20 individual exposures were co-added for each of the channels.  Since the LiF~1 channel provided the spectrum with the best signal to noise ratio, it was the only one used for our analysis of \ion{O}{6}.  As a consistency check, the \ion{O}{6} absorption recorded in the LiF~2 channel was also analyzed, yielding similar results.  The spectra were rebinned by a factor of three, after which the signal to noise ratio in the continuum near 1032 \AA\ was $\sim$~20 per rebinned sample.  The rebinning increased the pixel size near 1032 \AA\ from $\sim$~7 m\AA\ to $\sim$~20 m\AA, corresponding to a velocity sampling interval of $\sim$~5.9 km~s$^{-1}$.  A portion of the (rebinned) LiF~1 spectrum for HD~177989, from 1020 to 1045 \AA, is presented in Figure 1.

The \ion{O}{6} $\lambda$1037.62 absorption line is blended with strong absorption by \ion{C}{2} and H$_2$, and hence only the \ion{O}{6} $\lambda$1031.93 line was  measurable.  The velocity calibration in this region was performed by fitting the H$_2$ 6--0 P(3) and 6--0 R(4) lines at 1031.19 and 1032.36 \AA\ with Gaussians and measuring their central velocities.  Determining the velocity registration of the H$_2$ lines was complicated in that structure near 0 km~s$^{-1}$ is apparent in atomic low ion profiles produced by higher resolution studies (SSH; Sembach et al. 1993), but is unresolved by \emph{FUSE} ($\lambda/\Delta\lambda \sim 11,500$, or 26 km~s$^{-1}$; see below).  The H$_2$ lines were registered to an LSR velocity of 4.4 km~s$^{-1}$, the column density weighted average velocity of \ion{Cl}{1} $\lambda$1347.24 measured by STIS ($\lambda/\Delta\lambda \sim 110,000$, or FWHM $\sim$~2.7 km~s$^{-1}$), which has a better determined wavelength solution than \emph{FUSE}.  H$_2$ and \ion{Cl}{1} should coexist in the same interstellar regions since \ion{Cl}{1} is created through a rapid charge-exchange reaction between \ion{Cl}{2} and H$_2$ (Dalgarno et al. 1974; Jura \& York 1978).  This procedure introduces an absolute velocity uncertainty of $\sim$~5 km~s$^{-1}$ into the \ion{O}{6} spectrum.

The \ion{O}{6} absorption is contaminated by HD 6--0 R(0) $\lambda$1031.91, which is at $-4.1$ km~s$^{-1}$ relative to the \ion{O}{6} rest wavelength of 1031.93 \AA.  Since HD is expected to exist only in cold gas, its small \emph{b}-values (1--3 km~s$^{-1}$) and large optical depths place the lines on the flat part of the curve of growth (COG), where the absorption is only weakly dependent on differences in oscillator strength.  This is evidenced by the nearly identical equivalent widths of the HD 8--0 R(0) and 5--0 R(0) lines at 1011.46 and 1042.85 \AA, which have line strengths, \emph{f}$\lambda$, factors of 1.05 higher and 1.12 lower, respectively, than the HD line at 1031.91 \AA\ (Sembach 1999).  Hence, the Gaussian component fits to the 8--0 R(0) and 5--0 R(0) lines were averaged together to approximate the 6--0 R(0) line, and then were divided out of the \ion{O}{6} profile.  The H$_2$ wavelengths used are from the list compiled by Barnstedt et al. (2000), the HD line strengths are from Sembach (1999), and data for the atomic transitions are from Morton (1991).

The spectral resolution in the LiF~1 channel near \ion{O}{6} was approximated by plotting lines of molecular hydrogen on a model COG, and then finding the best-fit Doppler spread parameter $b$.  The proximity of the H$_2$ 6--0 P(3) and 6--0 R(4) lines to \ion{O}{6} $\lambda$1031.93 provided an impetus to use lines of these rotational quantum numbers to measure the $b$-value.  However, the detected $J = 4$ lines provided too small a range of \emph{f}-values to produce a good estimate of $b$, so the $J = 3$ lines were used.  Plotting the equivalent widths of these lines on a model COG gave $b = 6.5^{+3.5}_{-1.3}$.  Assuming a single Gaussian instrumental function, the H$_2$ lines within 15 \AA\ of \ion{O}{6} were fitted with Gaussian components having $b = 6.5$ km~s$^{-1}$, and the instrumental resolution was adjusted until the $\chi^{2}$ of the fit was minimized.  The mean instrumental resolution found by this method was $R = \lambda/\Delta\lambda \sim 11,500$, or $\sim$~26 km~s$^{-1}$ FWHM.  This value of $R$ was adopted in the component fits discussed in \S4.2.  However, we remark that, due to unresolved substructure in the H$_2$ profiles, and the possibility of a Doppler spread parameter larger than 6.5 km~s$^{-1}$, the actual instrumental profile width may in fact be slightly less than 26 km~s$^{-1}$.  Nonetheless, the component widths of \ion{O}{6} listed in Table 1 scarcely change even when $R$ is increased to 15,000, (20 km~s$^{-1}$ FWHM), the nominal resolution of most \emph{FUSE} data.

We normalized the absorption profiles by fitting low order Legendre polynomials to the continua. Figure 2 illustrates the continuum placement for \ion{O}{6} $\lambda 1031.93$, as well as the model of the contaminating HD $\lambda$1031.19 line.  Weak absorption by \ion{Cl}{1} at 1031.51 \AA\ influenced the continuum of the \ion{O}{6} line near $-120$ km~s$^{-1}$, but had negligible effect on the strength of the Scutum Supershell absorption at +42 km~s$^{-1}$.  The function represented by the dashed line in Figure 2 was divided out of the spectrum to produce the \ion{O}{6} continuum normalized profile depicted in Figure 3.

\section{OBSERVATIONAL RESULTS}

\subsection{Kinematics}

Figure 3 contains continuum normalized profiles for various interstellar absorption lines in the spectrum of HD~177989.  The profiles of \ion{S}{2} $\lambda$1253.81, \ion{Si}{4} $\lambda$1393.76, \ion{C}{4} $\lambda$1550.77, and \ion{N}{5} $\lambda$1238.82 are from observations obtained by the STIS, while \ion{Fe}{2} $\lambda$2374.46 was observed by the GHRS.  These profiles also appear in SSH, and have resolutions of approximately 3 km~s$^{-1}$ FWHM.  The \ion{Fe}{2} $\lambda$1144.94, \ion{P}{2} $\lambda$1152.82, and \ion{O}{6} $\lambda$1031.93 profiles are from the \emph{FUSE} LiF~1 channel.

The top half of Figure 3 illustrates the kinematic properties of low ion absorption along the line of sight to HD~177989.  The ions observed by the STIS reveal strong Local Arm absorption near 0 km~s$^{-1}$, while weaker features are seen near +36 and +56 km~s$^{-1}$.  The \ion{Fe}{2} and \ion{P}{2} lines from \emph{FUSE} show the same general characteristics, albeit at a lower resolution.  The features at +36 and +56 km~s$^{-1}$ in particular are blurred to the point that it is difficult to discern them from each other.  Due to that complication, no further analysis was performed with the low ion data from the \emph{FUSE} spectra.

The absorption by the more highly ionized species at the bottom of Figure 3 exhibits significantly different properties than the low ion absorption.  \ion{Si}{4} and \ion{C}{4} display weak absorption near $-50$ and $-13$ km~s$^{-1}$, and very strong absorption near +18 and +42 km~s$^{-1}$ (see SSH).  There is no known Galactic structure that is expected to produce absorption at $-50$ km~s$^{-1}$ in this direction.  The $-13$ km~s$^{-1}$ component may occur in an interface region between the hot Local Bubble gas and the surrounding neutral cloud.  The features at +18 and +42 km~s$^{-1}$ are due to gas lying $\sim$~0.4 and $\sim$~0.6--0.8 kpc below the Sagittarius and Scutum spiral arms, respectively (Callaway et al. 2000).  Note that the +42 km~s$^{-1}$ component is intermediate in velocity to the double absorption present in the low ion profiles, suggesting that the low ions dwell in cooler gas bounding the Scutum Supershell, while the higher ions reside in the hot inner regions.  The absorption profile for \ion{N}{5} is much less structured than, but consistent with, those of \ion{Si}{4} and \ion{C}{4}, with the exception that an additional component appears to be present at $\sim$~3 km~s$^{-1}$.  It is possible that this absorption is related to the Local Bubble/neutral cloud interface 130 pc distant in this direction (Crutcher \& Lien 1984).

The \ion{O}{6} profile is similar to \ion{N}{5} in that its component structure is much less pronounced than for \ion{C}{4} and \ion{Si}{4} (although this is due in part to the resolution of \emph{FUSE}).  The velocity range spanned by the \ion{O}{6} absorption, from $-70$ to +80 km~s$^{-1}$, is compatible with that for the other high ions.

\subsection{High Ion Component Structure}

Since the different absorption features in the high ion profiles overlap in velocity, we characterize the absorption by Gaussian component fits to provide more detailed information about the sight line velocity structure.  The fitting procedure follows that of Sembach et al. (1993) and assumes a single Gaussian instrumental spread function of FWHM 2.6 km~s$^{-1}$ for \ion{Si}{4}, \ion{C}{4}, and \ion{N}{5} (from STIS), and 26 km~s$^{-1}$ for \ion{O}{6}.  The fits for these four ions are portrayed in Figure 4, and the values of the resulting central velocity $<v_{\rm i}>$, Doppler spread parameter $b_{\rm i}$, and column density $N_{\rm i}$ for each component, as well as the integrated column density $N_{\rm tot}$, are presented in Table 1.  The errors for these quantities account for continuum placement, as well as statistical uncertainties in both the fit and the data itself, but do not account for errors in assumed velocity structure.

The fits for \ion{Si}{4}, \ion{C}{4}, and \ion{N}{5} are from SSH.  Both doublet lines for these three ions were fitted simultaneously to give the values in Table 1, although only the indicated line is shown in Figure 4.  Simultaneous doublet fitting was not possible for \ion{O}{6}, since the $\lambda$1037.62 line was thoroughly blended with strong H$_2$, \ion{C}{2}, and \ion{C}{2}$^*$ absorption.  Both \ion{C}{4} and \ion{Si}{4} required four components, which are labeled 1--4 in Table 1, from negative to positive velocities.  Since the \ion{N}{5} absorption has a significantly lower signal to noise, its components were fixed to be at the same velocities as those of \ion{C}{4}.  In addition, a narrow ($b \sim$ 7 km~s$^{-1}$) feature was needed at +2.6 km~s$^{-1}$ to produce an adequate fit to the \ion{N}{5} profile.  The column densities derived for the various components based on the 2.6 km~s$^{-1}$ resolution STIS measurements of \ion{Si}{4} and \ion{C}{4} should be quite reliable because the intrinsic absorption should be fully resolved, regardless of whether the gas is collisionally ionized or photoionized.  Furthermore, the simultaneous fitting of both members of the respective doublet lines having \emph{f}-values differing by a factor of two permitted a reliable assessment of the degree of saturation of the absorption (see SSH).  The \ion{N}{5} component column densities are less reliable since it was necessary to assume a kinematic model based on the \ion{Si}{4} and \ion{C}{4} measurements to perform the fitting.

The multiple component structure in the \ion{O}{6} absorption is not as clear as for \ion{Si}{4} and \ion{C}{4}.  This is probably due to a combination of the lower resolution of the \emph{FUSE} observations (FWHM $\approx$ 26 km~s$^{-1}$) and the intrinsically broader absorption produced by \ion{O}{6}.  Individual \ion{O}{6} velocity components are likely intrinsically broad since the thermal Doppler contribution to the broadening of \ion{O}{6} for gas at $T = 3 \times 10^5$ K corresponds to $b = 17.7$ km~s$^{-1}$ or FWHM = 29.5 km~s$^{-1}$, which is somewhat broader than the \emph{FUSE} resolution.  This implies that the \ion{O}{6} absorption should be nearly resolved by \emph{FUSE} and is unlikely to contain narrow unresolved saturated structure.  This is important since only the stronger doublet component of \ion{O}{6} is available for analysis.

The principal difficulty in the analysis of the \ion{O}{6} profile is separating the blended absorption from the different absorbing structures along the line of sight.  To perform the separation, the \ion{O}{6} component fits were constrained to have the same velocities as components 1--4 seen in \ion{C}{4}.  The resulting fit reveals three very broad components, with component 1 of the \ion{C}{4} model absent (although the absence of this feature is likely an effect of resolution; see the discussion below).  Since \ion{O}{6} is expected to resemble the \ion{N}{5} absorption more than \ion{C}{4} (as neither is likely to be photoionized by starlight, whereas \ion{C}{4} can), it is notable that its profile contains two fewer features.  The resolution of \emph{FUSE} is a likely culprit for the discrepancy with the five-component structure of \ion{N}{5}, as the narrow and overlapping features are effectively washed out of the \ion{O}{6} spectrum.  In fact, when the \ion{N}{5} profile as observed by the STIS is convolved with a 26 km~s$^{-1}$ FWHM Gaussian, the narrow feature at +2.6 km~s$^{-1}$ is no longer visible, and the rest of the absorption is similar to that seen in the \emph{FUSE} spectrum of the \ion{O}{6} line, although the \ion{O}{6} profile extends to larger positive velocity.  We conclude that \ion{O}{6} absorption at $-50$ and +3 km~s$^{-1}$ may indeed be present, but overlaps in velocity with neighboring components enough so as to be indistinguishable, given the \emph{FUSE} instrumental resolution.

A notable attribute of the \ion{O}{6} absorption is the broad widths of the components, which have $b$-values more than twice as large as for the other high ions.  The unresolved structure near $-50$ and +3 km~s$^{-1}$ will act to extend the wings and thereby broaden the two Gaussians at $-13$ and +18 km~s$^{-1}$.  However, additional unresolved structure is not expected near the Scutum peak, and hence the width for that component appears to be better constrained.  It is improbable that the line is broadened purely by thermal means, since the \emph{b}-value of 32.0 $\pm$ 2.4 for component 4 would imply a temperature of nearly 10$^6$~K.  \ion{O}{6} is a poor tracer of gas at such high temperatures, due to its low fractional abundance.  The breadth of this line results from the extension of the absorption to somewhat higher positive velocities than for the other species.  We note that the error in the zero-point velocity registration of the \ion{O}{6} spectrum, $\pm5$ km s$^{-1}$, could reduce the \emph{b}-value of the Scutum component by $\sim$~6 km s$^{-1}$.  Nevertheless, even this decreased value of $b \sim 26$ km s$^{-1}$ is considerably larger than is observed in the spectra of the other high ion species.  It is reasonable to question the effect lower resolution may have on the other high ion species.  A free fit of the blurred \ion{N}{5} profile mentioned above produced a component near +42 km s$^{-1}$ with parameters similar to the Scutum component in the higher resolution STIS data.  Specifically, this fit yielded a narrow ($b \sim 13$ km s$^{-1}$) width for the Scutum feature.  It appears that the \ion{O}{6} samples gas in a more kinematically disturbed region than the high ions observed by the STIS.  This is discussed in more depth in the following section.

It should be noted that the fitting model produced for these ions is not unique, since the actual absorption may not conform with the Gaussian models of optical depth that are assumed.  In the case of \ion{O}{6}, however, the absorption related to the Scutum Supershell is reproduced in a number of different fits.  A free fit of the \ion{O}{6} profile, for which $<v_{\rm i}>$, $b_{\rm i}$, and $N_{\rm i}$ were allowed to vary independently, revealed a three component structure with the parameters ($<v_{\rm i}>$ [km~s$^{-1}$], $b_{\rm i}$ [km~s$^{-1}$], $N_{\rm i}$ [cm$^{-2}$]) equalling ($-18.0, 36.2, 6.31 \times 10^{13}$), ($7.8, 4.1, 2.24 \times 10^{13}$), and ($41.5, 26.9, 8.91 \times 10^{13}$) for the respective components.  The non-uniqueness of the model fit listed in Table 1 is illustrated by the differing parameters of the components of the free fit at -18.0 and 7.8 km~s$^{-1}$.  However, the feature at +41.5 km~s$^{-1}$ (near the supershell velocity) in the free fit corresponds well with component 4 in Table 1. The reproducibility of the Scutum absorption indicates that the parameters found for it are reasonably well-defined.

\section{ANALYSIS}

The high ion absorption along the line of sight to HD~177989, in particular that associated with the Scutum Supershell, divulges intriguing characteristics that deviate from most Galactic interstellar environments.  Figure 4 illustrates the model component fits of the absorption, with the numerical results presented in Table 1.  The column densities in the Scutum region (+42 km~s$^{-1}$, labeled Component 4 in Table 1) are N(\ion{Si}{4}) = $(3.59 \pm 0.09) \times 10^{13}$, N(\ion{C}{4}) = $(1.78 \pm 0.04) \times 10^{14}$, N(\ion{N}{5}) = $(8.89 \pm 0.79) \times 10^{12}$, and N(\ion{O}{6}) = $(7.76 \pm 0.75) \times 10^{13}$ cm$^{-2}$.  Such large quantities of these high ions imply a sizeable contribution from the supershell.

A noteworthy trait that can be discerned from Figure 4 is the large width of the \ion{O}{6} absorption compared to the other high ions.  This indicates that the \ion{O}{6} traces gas that is not entirely coincident with the gas containing \ion{Si}{4}, \ion{C}{4}, and \ion{N}{5}.  However, the agreement of central velocities in free fits of \ion{O}{6} suggests a physical association of the \ion{O}{6} and other high ions.  Mac Low et al. (1989) modeled the dynamics of superbubble blowout, and found that blowout initiates at the polar cap of the shell, where the ambient density is lowest.  The shell then fragments and hot gas from the interior is ejected into the halo.  It is possible that the absorption from \ion{C}{4}, \ion{Si}{4}, and \ion{N}{5} originates in shell fragments above the center of the shell, while \ion{O}{6} samples hot gas that is permeating the high latitude cloud.  A component of the flow parallel to the disk lends to the broad wings of the line.  The conformity of central velocities in this picture is thus a result of the hot gas flowing above the polar region of the shell, past the fragments containing the other high ions.  Although it might seem unlikely that other species do not also trace this outflowing gas, such a phenomenon has been detected in other directions.  The sight lines to several quasars reveal extended Galactic absorption in \ion{O}{6} that is not seen in ions with lower ionization potential.  A particular case is the line of sight to 3C~273 (Sembach et al. 1997; Sembach et al. 2001), which passes through the edges of Radio Loops I and IV.  These are also superbubbles suspected to have been created by a series of supernova explosions.  A positive velocity wing is seen in \ion{O}{6} that is not evident in other absorption lines, which Sembach et al. attribute to hot gas being vented into the halo from the disk.  It is possible, both in this and the HD~177989 sight line, that other observed ions are present in the hot gas, but low fractional abundances probably place them below the detection limit.

As was noted by SSH, the Scutum component contains rather atypical column density ratios.  These are listed in Table 2 for the supershell, as well as for the total integrated sight line.  The values of N(\ion{C}{4})/N(\ion{Si}{4}) = 4.96 $\pm$ 0.17 and N(\ion{C}{4})/N(\ion{N}{5}) = 20.0 $\pm$ 1.8, when compared to Galactic averages of N(\ion{C}{4})/N(\ion{Si}{4}) = 4.3 $\pm$ 1.9 and N(\ion{C}{4})/N(\ion{N}{5}) = 4.6 $\pm$ 2.4 (Sembach et al. 1997; Savage et al. 1997), indicate that \ion{C}{4} and \ion{Si}{4} are enriched in this region, without a corresponding enhancement in \ion{N}{5}.  The \emph{FUSE} observations give ratios of N(\ion{C}{4})/N(\ion{O}{6}) = 2.29 $\pm$ 0.23 and N(\ion{O}{6})/N({\ion{N}{5}) = $8.73 \pm 1.15$.  Most hot gas models have difficulty in reproducing such ratios.  For example, it is shown in Table 10 of Sembach et al. (1997) that models such as cooling Galactic fountains (Shapiro \& Benjamin 1992), magnetized thermal conduction fronts (Borkowski, Balbus, \& Fristrom 1990), and supernova remnant bubbles (Slavin \& Cox 1992) predict N(\ion{C}{4})/N(\ion{N}{5}) $\sim$~1.8--6.8 and N(\ion{C}{4})/N(\ion{O}{6}) $\sim$~0.1--0.5, which are far lower than those observed along the HD~177989 sight line.  The only hot gas model that produces such abundant quantities of \ion{C}{4} relative to other high ions is that of turbulent mixing layers (TMLs; Begelman \& Fabian 1990; Slavin, Shull \& Begelman 1993).  In a TML, hot ($T \sim$ 10$^6$ K) and cool ($T \sim$ 10$^2$--10$^4$ K) gas mix in the presence of shear flows; the highly ionized atoms are created in the turbulent interface regions.  The mixing layers could be formed when hot gas from the interior of the superbubble flows past cooler shell fragments on its journey into the halo.  Such an explanation is also invoked by Heckman et al. (2001) for a blowout region in the dwarf starburst galaxy NGC~1705.  SSH propose a TML model for the Scutum Supershell with a post-mixed gas temperature of $T \sim 2 \times 10^5$ K and mixing velocity of $v_t \approx$ 65 km~s$^{-1}$.  This combination of physical parameters predicts ionic ratios of N(\ion{C}{4})/N(\ion{Si}{4}) $\sim$~3.3, N(\ion{C}{4})/N(\ion{N}{5}) $\sim$~20, N(\ion{C}{4})/N(\ion{O}{6}) $\sim$~2.5, and N(\ion{O}{6})/N(\ion{N}{5}) $\sim$~8.  In order to produce the observed column densities, roughly 150--200 mixing layers are necessary.  This requires a highly fragmented medium which, according to Mac Low et al. (1989), is expected in a region of superbubble blowout.  Since the \ion{O}{6} also traces hot gas unrelated to the TML region (that is, gas permeating the high latitude cloud), it is not surprising that the \ion{O}{6} column density is slightly under-produced by the model.  The ionic ratios involving \ion{C}{4} and \ion{Si}{4} could be brought further into accord with the observations if depletion or photoionization (discussed later in this section) is accounted for.

It is significant that TMLs do in fact yield ionic ratios in fair agreement with the observations, but inspection of the component widths in Table 1 shows that the line width of \ion{C}{4} disallows a temperature over T = 1.54 $\times 10^5$ K, compared to the temperature of 2 $\times 10^5$ K for the post-mixed gas in the model.  This maximum temperature is calculated by assuming pure thermal Doppler broadening.  We note that the errors from the STIS data include only the statistical uncertainties from a $\chi^2$ test of the fit, and do not reflect the dependence of the \emph{b}-value uncertainty with that for the central velocity.  When this systematic error is taken into consideration, the uncertainty in the Doppler spread parameter is much larger, on the order of a few km~s$^{-1}$.  The temperature of the post-mixed gas in the TML is thus attainable within the errors of the observed line widths.  Nevertheless, only a small portion of the line widths, with the exception of \ion{O}{6}, can be attributed to turbulent motion in the mixing layers.  This remains a serious caveat to the proposed TML model.

It is sensible to suspect that photoionization may help to produce the copious number of \ion{C}{4} and \ion{Si}{4} ions seen in this region.  Certainly, the fragmented top of the shell will allow stellar photons from the disk to penetrate into the high latitude realm that is sampled in the sight line.  We remark that such a process may indeed produce some of the observed \ion{Si}{4} column, which is slightly under-produced in the TML model, but is unlikely to account for all of the \ion{Si}{4} and \ion{C}{4} absorption.  The main problem with photoionization is that the models yield a ratio of N(\ion{C}{4})/N(\ion{Si}{4}) that is less than unity except in the most rarefied environs (Bregman \& Harrington 1986), compared to the value of 4.96 $\pm$ 0.17 that is observed.  Depletion of silicon into dust could increase the predicted ratio somewhat, but silicon is depleted only by $\sim$~$-0.25$ dex in the warm neutral medium of the halo (Savage \& Sembach 1996).  This value is similar to that found for carbon in diffuse clouds in the disk (Sofia et al. 1997).  Thus, photoionization does not appear to be the dominant process in the production of \ion{C}{4} and \ion{Si}{4} in the Scutum Supershell region.

In summary, it appears that the high ions are created predominantly in turbulent mixing layers above the polar region of the shell.  This explanation makes physical sense, as outflowing hot gas produces the required shear flow past the cooler fragmented shell during blowout.  The very large observed width of the \ion{O}{6} absorption can be attributed to the diffusion of hot gas into the high latitude cloud, although much of the column density is produced in the TMLs.  The ionic ratios predicted by the TML models can be brought into even better agreement with the observations if a portion of the \ion{Si}{4} column is generated by photoionization.

Observations of other sight lines through this region could be used to verify our proposed explanation.  In their multi-frequency study of the Scutum Supershell, Callaway et al. (2000) presented IUE spectra of \ion{Si}{4} and \ion{C}{4} for the stars HD~175876, HD~175754 and HD~177989.  These results were used to probe gas in the shell and provide limits on the distance to the region, but the resolution of the IUE did not permit the column density of the Scutum component to be measured in these sight lines.  Unfortunately, suitably distant UV-bright probe stars are very difficult to find, and other such observations that actually probe the gas in the supershell do not exist at this time.

\section{COMPARISON OF THE HD~177989 AND 3C 273 SIGHT LINES}

The only other line of sight through a Galactic supershell that has been observed with the \emph{FUSE} satellite is that for the QSO 3C~273, which lies in the direction of Radio Loops I and IV.  Specifically, this direction ($l = 290.0^{\rm o}$, $b = +64.4^{\rm o}$) passes through the edge of Loop~IV, and lies near the edge of Loop~I and the North Polar Spur.  These loops are thought to have been formed from a series of supernova explosions that have interacted with the surrounding interstellar medium.

Intermediate resolution (FWHM $\sim$~20 km~s$^{-1}$) GHRS observations of this sight line (Sembach et al. 1997) reveal properties of high ion absorption by \ion{C}{4}, \ion{Si}{4}, and \ion{N}{5}, while \emph{FUSE} observations (Sembach et al. 2001) provide information of the \ion{O}{6} absorption.  The column densities of these species are very large:  log~N(\ion{Si}{4}) = 13.78 $\pm$ 0.04, log~N(\ion{C}{4}) = 14.49 $\pm$ 0.03, log~N(\ion{N}{5}) = 13.87 $\pm$ 0.06, and log~N(\ion{O}{6}) = 14.73 $\pm$ 0.04.  Interestingly, although \ion{C}{4} and \ion{Si}{4} are certainly enhanced in this region, it is clear that the number of \ion{N}{5} and \ion{O}{6} ions are also magnified, much more so than in the Scutum Supershell.  The corresponding ionic ratios are tabulated in the fourth column of Table 2.  The ratios N(\ion{C}{4})/N(\ion{Si}{4}) = 4.8 $\pm$ 0.5 and especially N(\ion{C}{4})/N(\ion{N}{5}) = 4.0 $\pm$ 0.6 are closer to the Galactic averages than is seen in the Scutum absorption.  Furthermore, the value of N(\ion{C}{4})/N(\ion{O}{6}) = 0.59 $\pm$ 0.07 is only about one quarter as large as for HD~177989.  Different processes appear to be at work in the two regions.

The absorption profiles of the high ions toward 3C~273 are broad, indicating they are present in a highly turbulent medium.  The velocity range of absorption is roughly --100 to +100 km~s$^{-1}$.  From the column density ratios in Table 2, Sembach et al. (1997) propose that multiple processes are required to account for the properties seen.  The negative velocity absorption fits well with a model that consists of roughly equal contributions from TMLs and the magnetized thermally conductive interfaces studied by Borkowski et al. (1990).  Meanwhile, the positive velocity absorption differs considerably and displays attributes resembling those of a cooling Galactic fountain.  The \ion{O}{6} spectrum seems to corroborate this view, and also contains a very broad, shallow wing from +100 to +240 km~s$^{-1}$ that is not seen in any other species.  Sembach et al. (2001) believe this wing is likely to be associated with hot gas flowing from the disk to the halo.

It should be noted that other sight lines toward Radio Loops I and IV have been observed that display quite different properties than the 3C~273 direction.  In particular, Sembach et al. (1997) also presented high resolution GHRS spectra of the star HD~119608 ($l = 320.4^{\rm o}$, $b = 43.1^{\rm o}$), which lies toward the center of the loops.  Like 3C~273, the \ion{C}{4} column is large along this line of sight, although such an enhancement is \emph{not} reflected in \ion{Si}{4} or \ion{N}{5}.  The ionic ratios of N(\ion{C}{4})/N~\ion{Si}{4}) = 8.1 $\pm$ 1.1 and N(\ion{C}{4})/N(\ion{N}{5}) = 10.7 $\pm$ 2.1 exhibit this discrepancy.  The ionizing mechanisms are accredited to TMLs and magnetized thermal conduction fronts, as is the negative velocity absorption in 3C~273.  The absorption profiles seen in the spectrum of HD~119608 also are much narrower than those in the 3C~273 spectrum, indicating that the center of the shell is less turbulent than near its edges.  Unfortunately, severe stellar blending prevents a study of \ion{O}{6} absorption toward HD~119608 with \emph{FUSE}.

It is clear from the differences seen in the high ion profiles between HD~177989, HD~119608, and 3C~273 that the traits of absorption in a superbubble, and indeed the ionizing mechanisms themselves, are quite dependent on the actual position of the line of sight through the shell.  Despite such differences, some characteristics, such as the outflowing hot gas believed to be observed in both 3C~273 and HD~177989, are common to the sight lines.  Future observations will shed more light on the physical properties of superbubbles and their intra-shell variations.

\section{SUMMARY}

We now summarize the main results of our study.

1.  We have used the \emph{FUSE} satellite to analyze absorption in the direction of HD~177989 ($l = 17.8^{\rm o}$, $b = -11.9^{\rm o}$), a B0~III star that lies $\sim$~4.9 kpc distant.  This line of sight samples gas in the high latitude cloud above the Scutum Supershell (GS~018-04+44), which spans a diameter of $\sim 5^{\rm o}$ and lies $\sim$~3.5 kpc distant.  The \ion{O}{6} $\lambda$1031.93 absorption line has been related to STIS observations of \ion{Si}{4}, \ion{C}{4}, and \ion{N}{5} (SSH) in order to provide information about the ionizing mechanisms, and the processes by which hot gas from the interior of the supershell is ejected into the halo.

2.  Component fits of \ion{O}{6} reveal very broad absorption at the kinematic velocity of the Scutum Supershell.  The FWHM of this component is more than twice that of the features at the same velocity in the lines of \ion{Si}{4}, \ion{C}{4}, and \ion{N}{5}, and indicates that the \ion{O}{6} is not entirely coincident with these ions.  The breadth of the line probably results from \ion{O}{6} sampling hot gas from the supershell interior that is diffusing into the high latitude cloud, while the highly ionized atoms observed by STIS exist in a more confined, fragmented medium above the polar cap of the shell.

3.  The only hot gas model capable of reproducing the unusual ionic ratios N(\ion{C}{4})/N(\ion{Si}{4}) = 4.96 $\pm$ 0.17, N(\ion{C}{4})/N(\ion{N}{5}) = 20.0 $\pm$ 1.8, and N(\ion{C}{4})/N(\ion{O}{6}) = 2.29 $\pm$ 0.23 is that of turbulent mixing layers.  The mixing layers could be formed as hot gas from the interior of the bubble rushes past cooler material in shell fragments overlying the superbubble.  The 150--200 TMLs necessary to explain the column densities are consistent with such an inhomogeneous gas distribution.  A caveat to this model is that once thermal Doppler broadening is accounted for, the relatively narrow line widths of \ion{Si}{4}, \ion{C}{4}, and \ion{N}{5} allow little room for turbulent gas motion.

4.  The line of sight to HD~177989 has been compared to \emph{FUSE} and GHRS observations of absorption toward QSO~3C~273, which lies in the direction of the edges of Radio Loops I and IV.  The two sight lines reveal quite different high ion absorption, since \ion{N}{5} and \ion{O}{6}, as well as \ion{C}{4} and \ion{Si}{4}, appear to be enhanced in the Radio Loops.  The column density ratios are thus much closer to the Galactic averages than is observed in the Scutum high latitude cloud.  Meanwhile, GHRS observations of the star HD~119608, near the center of Loops I and IV, reveal properties that are more similar to those seen along the HD~177989 sight line.  It is apparent that the high ion absorption, as well as the ionizing mechanisms, can vary widely among sight lines through a superbubble.

This work was based on observations made with the NASA-CNES-CSA \emph{FUSE} mission operated by the Johns Hopkins University.  We thank the members of the \emph{FUSE} operations and science teams for providing this outstanding facility to the astronomical community.  We also thank Marilyn Meade who kindly reduced the \emph{FUSE} data that appears in this paper.  Financial support was provided through NASA contract NAS5-32985.

\clearpage

\begin{deluxetable}{ccccccc}
\tablecolumns{7}
\tablewidth{0pc} 
\tablecaption{High Ion Component Fits\tablenotemark{a}}
\tabletypesize{\small}
\tablehead{
\colhead{} & \colhead{} & \colhead{} & \colhead{$<v>$} & \colhead{$b$} & \colhead{$N$} & \colhead{$T_{max}$} \\
\colhead{Ion} & \colhead{$N_{\rm tot}$} & \colhead{Component} & \colhead{(km~s$^{-1}$)} & \colhead{(km~s$^{-1}$)} & \colhead{(cm$^{-2}$)} & \colhead{($10^{5}$ K)}}
\startdata
Si IV\tablenotemark{b}.... & 6.25 $\pm$ 0.12 (13) & 1 & -51.8 & 4.3$\pm$2.8 & 2.48$\pm$1.82 (11) & 0.31\\
 & & 2 & -13.7 & 11.5$\pm$0.6 & 6.52$\pm$0.37 (12) & 2.22\\
 & & 3 & +18.5 & 11.7$\pm$0.3 & 1.98$\pm$0.06 (13) & 2.30\\
 & & 4 & +41.7 & 12.7$\pm$0.3 & 3.59$\pm$0.09 (13) & 2.71\\
C IV\tablenotemark{b}..... & 2.94$\pm$0.53 (14) & 1 & -47.2 & 15.9$\pm$1.8 & 1.05$\pm$0.13 (13) & 1.82\\
 & & 2 & -12.9 & 13.4$\pm$0.6 & 3.22$\pm$0.17 (13) & 1.29\\
 & & 3 & +18.2 & 14.0$\pm$0.4 & 7.34$\pm$0.27 (13) & 1.41\\
 & & 4 & +42.4 & 14.6$\pm$0.2 & 1.78$\pm$0.04 (14) & 1.54\\
N V\tablenotemark{b, c}...... & 2.29$\pm$0.31 (13) & 1 & -47.2 & 9.1$\pm$6.4 & 1.35$\pm$1.09 (12) & 0.70\\
 & & 2 & -12.9 & 15.3$\pm$5.4 & 4.48$\pm$1.78 (12) & 1.97\\
 & & 3 & +18.2 & 12.1$\pm$3.9 & 4.93$\pm$1.75 (12) & 1.23\\
 & & 4 & +42.4 & 15.9$\pm$1.9 & 8.89$\pm$0.79 (12) & 2.13\\
 & & 5 & +2.6 & 7.0$\pm$2.3 & 3.29$\pm$1.28 (12) & 0.41\\
O VI\tablenotemark{d}..... & 1.82$\pm$0.18 (14) & 2\tablenotemark{e} & -12.9 & $44.2^{+1.1}_{-7.5}$ & $6.46^{+0.62}_{-1.44}$ (13) & 18.77\\
 & & 3\tablenotemark{e} & +18.2 & 30.9$\pm$3.6 & 3.98$\pm$1.27 (13) & 9.18\\
 & & 4 & +42.4 & 32.0$\pm$2.4 & 7.76$\pm$0.75 (13) & 9.84\\
\enddata
\scriptsize
\tablenotetext{a}{The resulting central LSR velocity, Doppler spread parameter, and column density is given for each component.  The actual fits are illustrated in Figure 4.  Absorption associated with the Scutum Supershell is at +42 km~s$^{-1}$ (component 4).  An upper limit to the temperature of the absorbing gas is given by assuming pure thermal Doppler broadening, $T_{max} = A[7.75b_{\rm i}$(km~s$^{-1}$)], where A is the atomic mass number of the ion.  The uncertainties given are 1-$\sigma$ estimates.}
\tablenotetext{b}{The values for Si~IV, C~IV, and N~V are taken from SSH, and were found via a simultaneous fit of both doublet lines for each ion.}
\tablenotetext{c}{N~V was fit by constraining its component velocities to be the same as those for C~IV, with the addition of a narrow feature (Component 5) at +2.6 km~s$^{-1}$.}
\tablenotetext{d}{A free fit of O~VI produced three components with the parameters ($<v_{\rm i}>$ [km~s$^{-1}$], $b_{\rm i}$ [km~s$^{-1}$], $N_{\rm i}$ [cm$^{-2}$]) equal to ($-18.0, 36.2, 6.31 \times 10^{13}$), ($7.8, 4.1, 2.24 \times 10^{13}$), and ($41.5, 26.9, 8.91 \times 10^{13}$).  The feature at +41.5 km~s$^{-1}$ in the free fit corresponds well to the Scutum Supershell absorption (component 4) of the model in the table.}
\tablenotetext{e}{Components 2 and 3 in O~VI probably contain unresolved structure from absorption near -50 and +3 km~s$^{-1}$.  This acts to broaden the lines and increase the column density.}
\end{deluxetable}

\clearpage

\begin{deluxetable}{cccc}
\tablecolumns{4}
\tablewidth{0pc} 
\tablecaption{High Ion Column Density Ratios\tablenotemark{a}}
\tablehead{
\colhead{} & \colhead{Scutum} & \colhead{} & \colhead{QSO} \\
\colhead{Ratio} & \colhead{Component} & \colhead{Total} & \colhead{3C~273\tablenotemark{b}}}
\startdata
$N$(\ion{C}{4})/$N$(\ion{Si}{4}) & 4.96 $\pm$ 0.17 & 4.71 $\pm$ 0.12 & 4.8 $\pm$ 0.5 \\
$N$(\ion{C}{4})/$N$(\ion{N}{5}) & 20.0 $\pm$ 1.8 & 12.8 $\pm$ 1.8 & 4.0 $\pm$ 0.6 \\
$N$(\ion{C}{4})/$N$(\ion{O}{6}) & 2.29 $\pm$ 0.23\tablenotemark{c} & 1.54 $\pm$ 0.32 & 0.59 $\pm$ 0.07 \\
$N$(\ion{O}{6})/$N$(\ion{N}{5}) & 8.73 $\pm$ 1.15\tablenotemark{c} & 8.34 $\pm$ 1.40 & 7.2 $\pm$ 1.1 \\
\enddata
\tablenotetext{a}{The column density ratios are given for the Scutum component, as well as for the total integrated profiles.  Errors are 1-$\sigma$ estimates.}
\tablenotetext{b}{The first two ratios for 3C 273 are taken from Table 10 of Sembach et al. (1997), while the last two are from Table 7 of Sembach et al. (2001).  \ion{O}{6} observations were obtained by the \emph{FUSE}, and those for \ion{Si}{4}, \ion{C}{4}, and \ion{N}{5} were procured from the GHRS.}
\tablenotetext{c}{The errors listed do not account for uncertainties of the velocity structure in \ion{O}{6}.  For example, the free fit parameters of the Scutum component (footnote \emph{d} of Table 1) yield ratios of $N$(\ion{C}{4})/$N$(\ion{O}{6}) = 2.00 and $N$(\ion{O}{6})/$N$(\ion{N}{5}) = 10.0.}
\end{deluxetable}

\clearpage

\figcaption[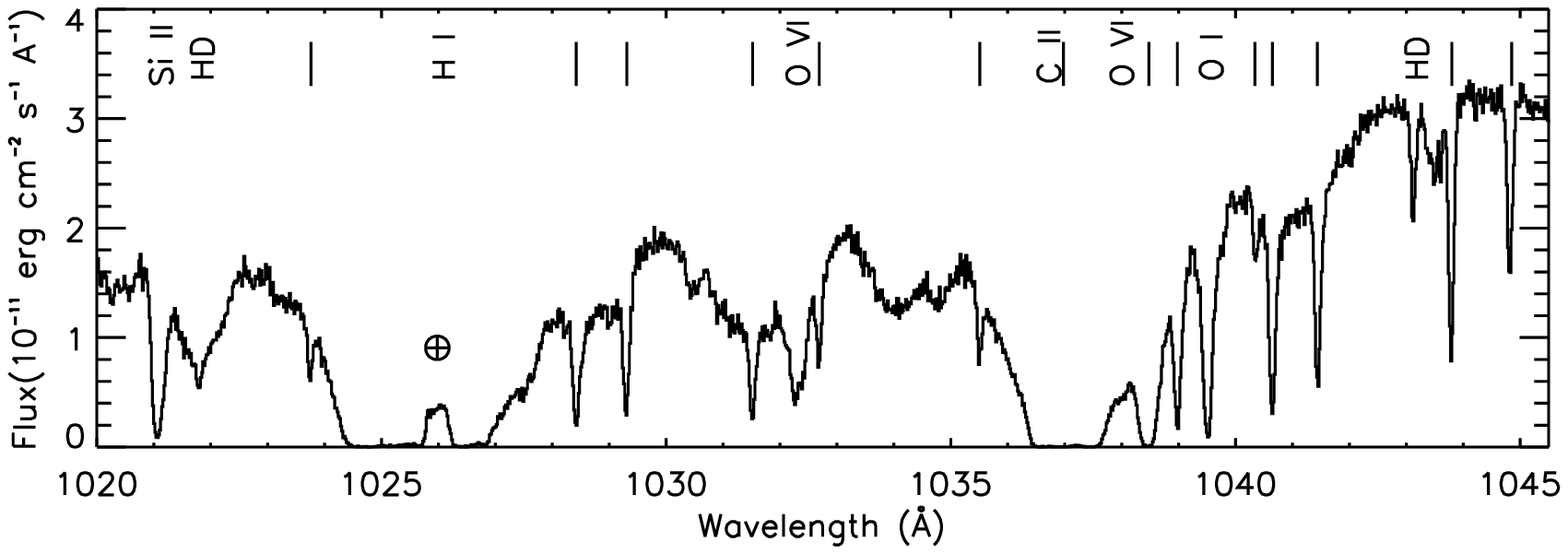]{A portion of the \emph{FUSE} LiF 1A spectrum for HD~177989 from 1020--1045 \AA\ is shown.  The data, uncorrected for wavelength offsets, have been rebinned by 3 pixels, corresponding to 0.02 \AA.  Interstellar absorption lines are identified above the spectrum; the vertical tick marks indicate H$_2$ lines.}

\figcaption[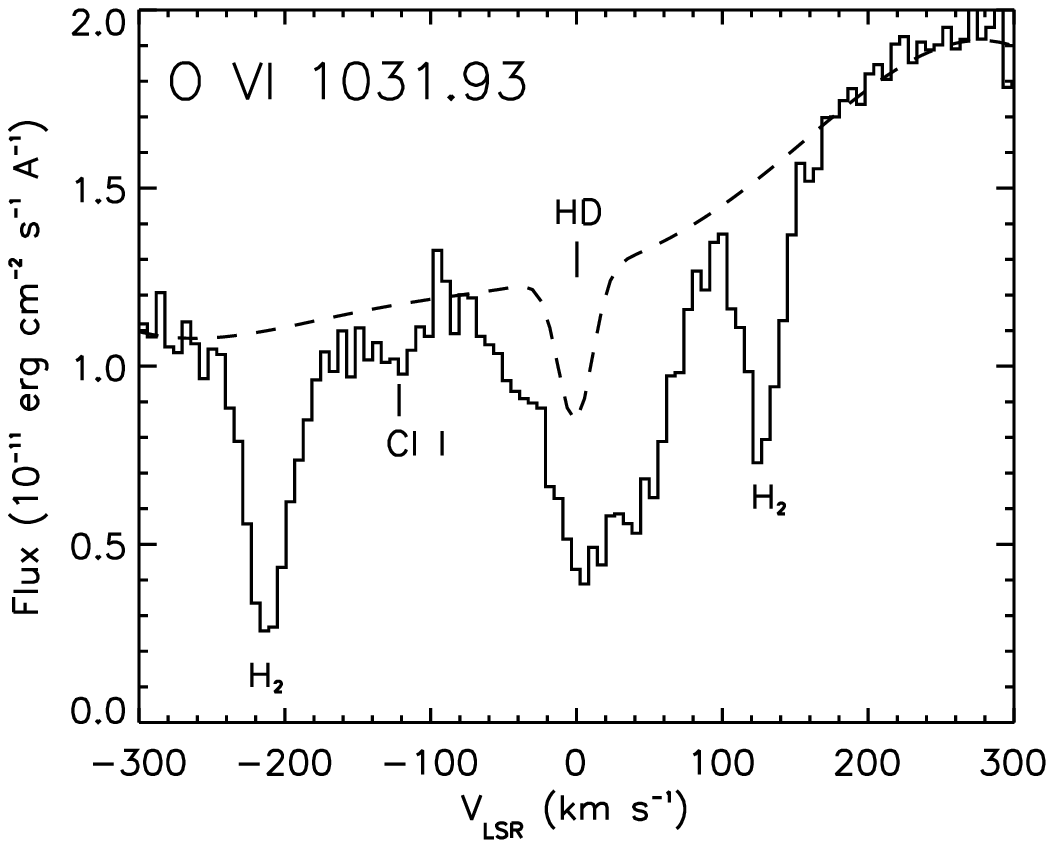]{\ion{O}{6} absorption at 1031.93 \AA\ in the HD~177989 spectrum is plotted in units of flux [$10^{-11}$ erg cm$^{-2}$ s$^{-1}$ \AA$^{-1}$] against LSR velocity in km~s$^{-1}$.  Other interstellar absorption lines present in the profile are indicated.  The dashed line illustrates the adopted continuum placement along with a model for the contaminating absorption by HD~6--0~R(0)~$\lambda$1031.91.  This model was produced by averaging the profiles of the similarly strong HD 8--0 R(0) and 5--0 R(0) lines at 1011.46 and 1042.85 \AA.  We also note that weak absorption by \ion{Cl}{1} at 1031.51 \AA\ affects the continuum placement of the \ion{O}{6} line near $-120$ km~s$^{-1}$.  The function represented by the dashed line was divided out of the spectrum to produce the continuum normalized \ion{O}{6} profile shown in Figure 3.}

\figcaption[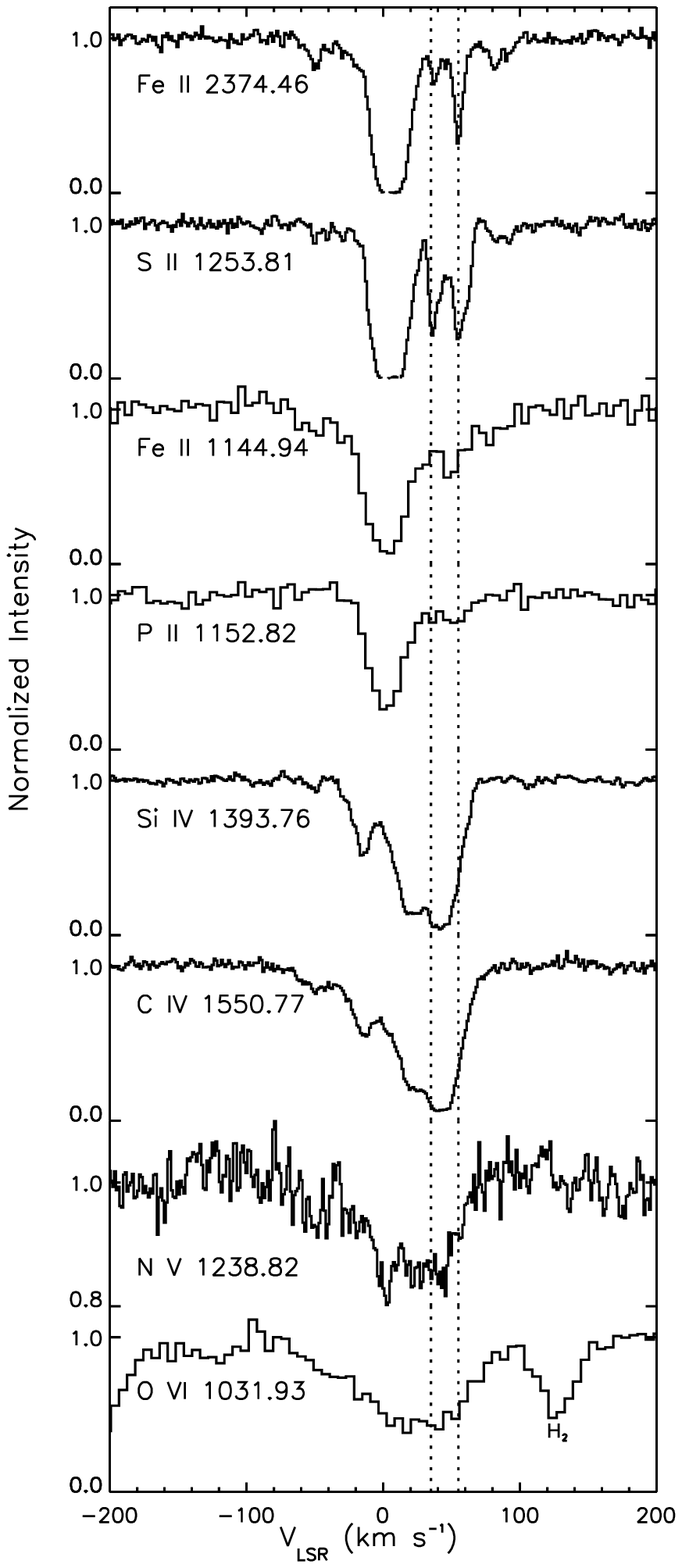]{Continuum normalized absorption profiles are plotted versus LSR velocity for the indicated species.  Five of the profiles also appear in SSH: observations of \ion{S}{2} $\lambda$1253.81, \ion{Si}{4} $\lambda$1393.76, \ion{C}{4} $\lambda$1550.77, and \ion{N}{5} $\lambda$1238.82 were obtained by the STIS, while \ion{Fe}{2} $\lambda$2374.46 was observed by the GHRS.  The other three profiles are from the \emph{FUSE} satellite.  The two dotted lines at +35 and +55 km~s$^{-1}$ correspond roughly to the velocity of Scutum Supershell absorption seen in \ion{Fe}{2} $\lambda$2374.46 and \ion{S}{2} $\lambda$1253.81.  The more highly ionized atoms in the supershell absorb at a velocity intermediate to the low ions ($\sim$~42 km~s$^{-1}$).  The scale of the \ion{N}{5} profile was enlarged to increase clarity.}

\figcaption[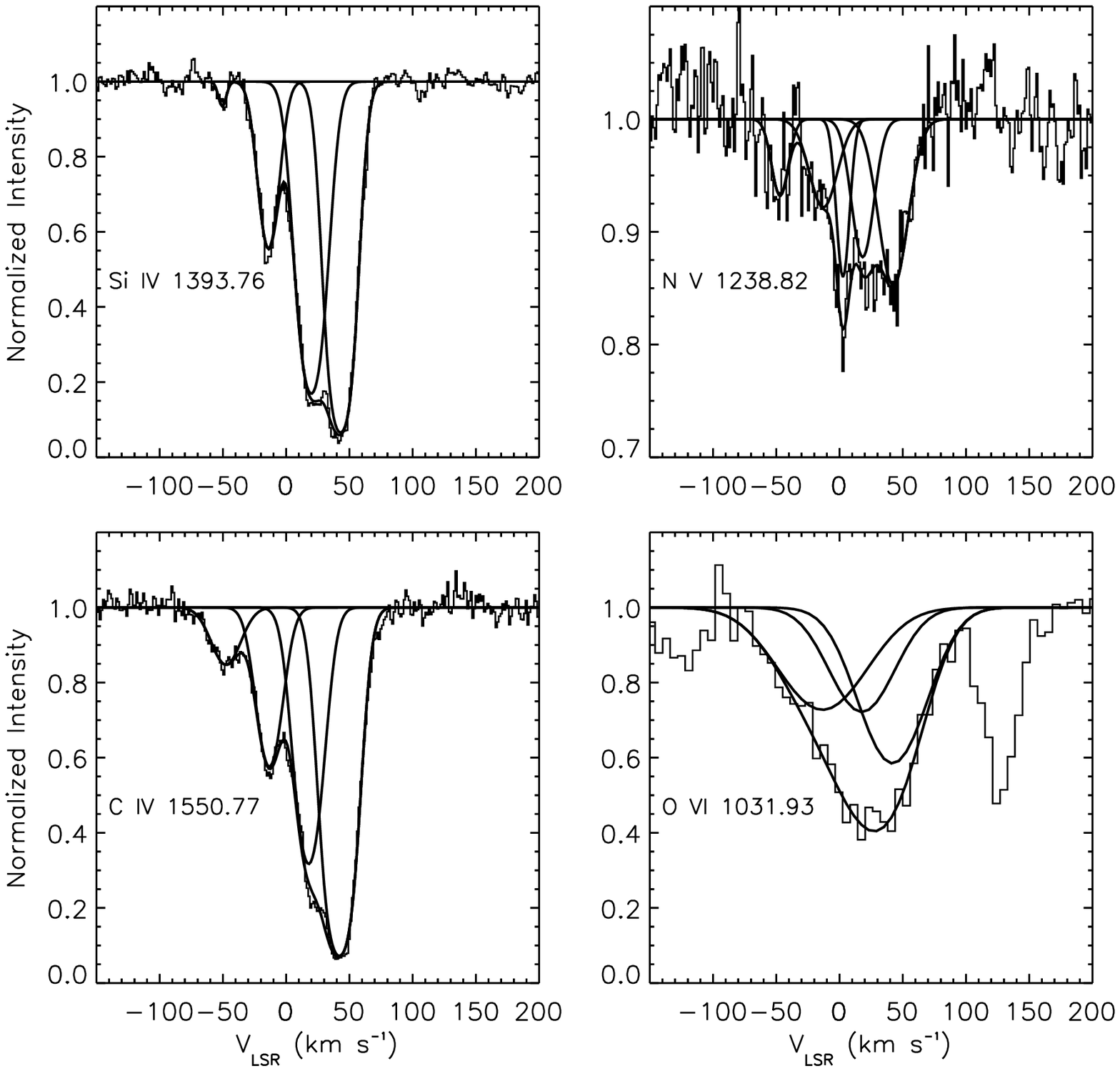]{Doppler broadened component fits to normalized intensity versus LSR velocity for \ion{Si}{4} $\lambda$1393.76, \ion{C}{4} $\lambda$1550.77, \ion{N}{5} $\lambda$1238.82, and \ion{O}{6} $\lambda$1031.93.  Both the individual components and the composite fit are shown.  The \ion{Si}{4}, \ion{C}{4}, and \ion{N}{5} fits were adapted from SSH.  \ion{Si}{4} and \ion{C}{4} were fit with four components, while \ion{N}{5} required five, and \ion{O}{6} only three.  Due to the weakness of the \ion{N}{5} absorption, its velocities were fixed to be the same as for \ion{C}{4}.  The same was done for \ion{O}{6}, although the component near -50 km~s$^{-1}$ was not required in the fit.  The H$_2$ R(4) 6--0 $\lambda$1032.36 line is included in the \ion{O}{6} component fit model, but is not plotted in the figure.  The Scutum Supershell absorption is at +42 km~s$^{-1}$.  We note that this component was similarly fit in a free fit of the \ion{O}{6} absorption profile, although the other two components differed.}

\clearpage

\begin{figure}
\plotone{fig1.ps}
\end{figure}

\clearpage

\begin{figure}
\plotone{fig2.ps}
\end{figure}

\clearpage

\begin{figure}
\plotone{fig3.ps}
\end{figure}

\clearpage

\begin{figure}
\plotone{fig4.ps}
\end{figure}

\end{document}